\documentclass[11pt,twoside]{article}

\usepackage{asp2006}

\usepackage[english]{babel}

\begin{document}

\title{Polarimetry from the Ground Up}
\author{C. U. Keller and F. Snik}
\affil{Sterrekundig Instituut Utrecht, PO Box 80000, NL-3508TA Utrecht, 
       The Netherlands}

\begin{abstract}
Ground-based solar polarimetry has made great progress over the last
decade. Nevertheless, polarimetry is still an afterthought in most
telescope and instrument designs, and most polarimeters are designed
based on experience and rules of thumb rather than using more formal
systems engineering approaches as is common in standard optical
design efforts. Here we present the first steps in creating a set of
systems engineering approaches to the design of polarimeters that
makes sure that the final telescope-instrument-polarimeter system is
more than the sum of its parts.
\end{abstract}

\section{Systems Engineering for Polarimetry}

Systems engineering ensures that the total is more than the sum of its
parts.  Systems engineering is essential for the successful design of
polarimeters where apparently unrelated effects such as instrumental
polarization due to oblique reflections in the telescope and detector
non-linearity couple to generate substantial systematic errors (Keller
1996).  When designing a polarimeter, it is therefore crucial to
optimize the entire system and not just the individual parts as has
often been done in the past.

While systems engineering is common in standard optics, it has been
largely absent in polarimetry because errors in polarization cannot
easily be expressed as a scalar and optical elements such as polarizers
and retarders have a major influence on the polarization. In the
following we will present initial thoughts on establishing systems
engineering approaches for polarimetry that will enable better
polarimeter designs with predictable performance.

Systems engineering for polarimetry models and helps us understand the
performance of polarimeter designs.  It must address at least the following
issues:
\begin{itemize}\itemsep=0pt
\item a definition of the polarimeter performance to quantitatively
  compare different polarimeter designs with ideal components;
\item a polarization error budget that can predict the performance
  based on known error statistics of real components;
\item methods to maximize the performance of a polarimeter design.
\end{itemize}
While the first issue has been addressed in the literature,
error budgets, a crucial systems engineering tool, have been
lacking.  In the following sections we present some initial ideas on
systems engineering approaches for polarimeter designs.

\section{Errors}

Errors in polarization measurements fall into two classes: statistical
(random) errors and systematic (instrumental) errors.  Systems
engineering must provide approaches to quantify and balance these two,
very different, error sources.

\subsection{Statistical Errors in Polarization Measurements}

The influence of statistical errors on the final polarization
measurement can be calculated using error propagation. These
calculations become relatively simple if the following assumptions
hold:
\begin{itemize}\itemsep=-1pt
\item there is a linear relation between the Stokes parameters of the 
      incoming light and the signals that are measured;
\item the noise does not depend on the position of a measurement
      in a sequence;
\item the noise is independent of the signal, which is the case if
      \begin{itemize}
      \item the noise is dominated by signal-independent detector noise 
            (e.g.\ read-out noise) or
      \item the light that is analyzed is only slightly polarized;
      \end{itemize}
\item the noise statistic has a Gaussian distribution.
\end{itemize}
For the large number of photons, which are needed for accurate
polarimetry, a Gaussian distribution is a good approximation to the
Poisson distribution of photon noise.  If the degree of polarization
of the incoming beam is small, the measurements will all have very
similar light levels, thereby justifying the assumption of
independence of noise and signal.

\subsection{Signal Matrix}

The intensities measured by the detector can be combined into a signal
vector $\mathbf S$, which is related to the incoming Stokes vector,
$\mathbf I$ by the signal matrix $\sf X$ (sometimes also called
the synthesis matrix, e.g.\ Tyo 2002),
\begin{equation}
{\mathbf S}={\sf X}{\mathbf I}\;.
\label{equ:keller_signalmatrix}
\end{equation}
$\sf X$ is a 4 by $m$ matrix, where $m$ is the number of intensity
measurements that contribute to the polarization measurement.  For
example, $m=4$ for most systems that use liquid crystals, while $m=8$
for a rotating retarder.  $\sf X$ is a function of the free
parameters of the polarimeter design.  Since the polarimeter optics
can be described by Mueller matrices, each row of $\sf X$
corresponds to the first row of the Mueller matrix describing the
particular intensity measurement as a function of the incoming Stokes
vector.

To determine the Stokes vector $\mathbf I$ from the measurements $\mathbf
S$, $\sf X$ needs to be inverted.  With
\begin{equation}
\sf Y = \sf X^{-1}
\end{equation}
the standard deviations of the Stokes parameters, $\sigma_{I_i}$, are
given by
\begin{equation}
\sigma_{I_i}=\sqrt{\sum_{j=1}^m {\sf Y}_{ij}^2\sigma_{S_j}^2}\,,
\end{equation}
where $\sigma_{S_j}$ is the standard deviation of the intensity in
measurement $j$.  In many cases, the latter does not depend on the
measurement number $j$ and one can rewrite the equation as
\begin{equation}
\sigma_{I^\prime_i}=\sigma_S \sqrt{\sum_{j=1}^m {\sf Y}_{ij}^2}\,.
\label{equ:keller_ErrorProp}
\end{equation}
These errors can then subsequently be propagated into errors in the
degree of circular, linear polarization and angle of linear
polarization or polarization ellipse parameters, and then into, e.g.,
solar magnetic field parameters. The last step is important because
the relations between Stokes parameters and magnetic field parameters
or other physical quantities of interest are not linear, and one
should optimize a polarimeter design according to what one wants to
measure.

\subsection{Systematic or Instrumental Errors}

Because of the large photon flux from the Sun, solar polarimeters are
largely limited by systematic (instrumental) errors rather than by
statistical errors.  A list and discussion of instrumental error
sources can be found in Keller (2002).  Our goal here is not to
understand the individual error sources but to discuss ways to 'add'
their influence such that we can predict the performance of a
polarimeter under non-ideal conditions.

\section{Polarimetric Efficiency}

If one is able to translate the science requirements into requirements
for the measurement of the Stokes parameters, then the concept of a
{\em polarimetric efficiency} is useful. Any definition of the
polarimetric efficiency must have the following properties:
\begin{itemize}\itemsep=0pt
\item comparable between different polarimeter designs and measurement
      approaches;
\item larger values should correspond to better designs;
\item independent of the intensity throughput;
\item consist of 4 quantities (``Stokes efficiency'');
\item the theoretical maximum efficiency shall be 1.
\end{itemize}

The polarimetric efficiency with which the component $i$ of the Stokes
vector $(I,Q,U,V)^T$ is measured was defined by del Toro Iniesta and
Collados (2000) as
\begin{equation}
\epsilon_i = \left(m\sum_{j=1}^m {\sf Y}_{ij}^2 \right)^{-{1\over 2}}\;.
\label{equ:keller_OptPolEff}
\end{equation}
Note that this definition is independent of the number of measurements
that contribute to the measurement and fulfills all the requirements
listed above if $\sf X_{i1}=1$.

Since the polarimetric efficiency is independent of the intensity
throughput, it is important to not directly compare polarimetric
efficiencies of different designs but to also properly take into
account the throughput of the corresponding design.  For instance, a
polarizing beam-splitter makes use of all photons, while a regular
linear polarizer will not transmit more than 50\% of all photons.

\subsection{Analytic Optimization}

When optimizing the design of a polarimeter, it is convenient to
define a scalar function of the free design parameters such that the
maximum or minimum of this {\em merit function} corresponds to the
optimum polarimeter design.  The polarimetric efficiency provides a
well-defined {\em merit function} for optimizing a polarimeter design,
in particular when one demands a fixed ratio (often unity) between the
efficiencies for the different polarized Stokes parameters.  Various
merit functions can be found in the literature, but they are all
closely related to the polarimetric efficiency.

For certain cases of merit functions, one can derive equations for the
properties of the optimum polarimeter (e.g.\ del Toro Iniesta and
Collados, 2000; Tyo 2002).  In practice, one may often use numerical
optimization schemes, but the analytic approach does reveal some of the
basic properties of an {\em optimum} polarimeter design and allows us
to derive the {\em maximum performance}.  The {\em maximum
  performance} is the best performance of all polarimeter designs,
while the {\sl optimum performance} is the best performance that can
be achieved with a given polarimeter design.  For a polarimeter with
maximum performance, $\epsilon_I=1$ and $\epsilon_Q^2+\epsilon_U^2+
\epsilon_V^2=1$.

We extend Eq.~\ref{equ:keller_signalmatrix} into the basic measurement
equation
\begin{equation}
{\mathbf S}={\sf X}({\mathbf v}) {\mathbf I} + {\mathbf n}\;.
\label{equ:keller_OptSignal}
\end{equation}
by explicitly showing the dependence on the free design parameters
$\mathbf v$ and the addition of random, zero-mean
noise $\mathbf n$.  A minimum of $m=4$ measurements is required to
determine all four Stokes parameters.

To determine an estimate ${\mathbf I}^\prime$ of the Stokes vector $\mathbf I$
from the measurements $\mathbf S$, $\sf X$ needs to be inverted.  If
$\sf Y$ (sometimes called the analysis matrix or demodulation
matrix) is the inverse of $\sf X$, we obtain
\begin{equation}
{\mathbf I}^\prime = {\sf Y} {\mathbf S} = 
                {\sf Y} \left({\sf X}({\mathbf v}){\mathbf I} + {\mathbf n}\right) =
                {\sf Y} {\sf X}({\mathbf v}){\mathbf I} + {\sf Y}{\mathbf n}\;.
\label{equ:keller_OptMeas}
\end{equation}
Our goal is to choose ${\sf X}({\mathbf v})$ and $\sf Y$ such as to
minimize the difference between ${\mathbf I}^\prime$ and $\mathbf I$
given the standard deviations $\sigma_{S_j}$ of the measurement errors
$n_j$.  We tackle this problem in three steps: 1) derive an equation
for the optimum $\sf Y$ for a given ${\sf X}({\mathbf v})$; 2) derive
an optimum $\sf X$ for the optimum $\sf Y$; and 3) choose optimum $\sf
X$ and $\sf Y$ to obtain a polarimeter with maximum performance.

\subsubsection{Optimum synthesis matrix}

For a given ${\sf X}({\mathbf v})$, $\sf Y$ is not necessarily
unique. It is obvious that apart from fulfilling ${\sf Y}{\sf X} = {\sf
1}$, $\sf Y$ should minimize ${\sf Y}{\mathbf n}$, given the standard
deviations $\sigma_j$.  This can be written as a minimization problem,
the solution of which is given by the generalized inverse (e.g.\
Albert 1972)
\begin{equation}
{\sf Y} = \left({\sf X}^T {\sf X}\right)^{-1}{\sf X}^T\;.
\label{equ:OptInverse}
\end{equation}
Among all possible $\sf Y$ that fulfill ${\sf Y}{\sf X} = {\sf 1}$, the
generalized inverse minimizes the sum of squares of its rows.
Therefore, the generalized inverse is the optimum synthesis matrix
$\sf Y$, given a signal matrix $\sf X$.  Since ${\sf X}^T{\sf X}$ is
symmetric and positive definite, it can be inverted with standard
matrix inversion algorithms.

Realizing that the sum of squares of the rows of $\sf Y$ are the
diagonal elements of ${\sf Y}{\sf Y}^T$, the optimum polarimetric
efficiencies for a given signal matrix $\sf X$ are derived (after some
linear algebra) as
\begin{equation}
\epsilon_{{\rm opt},i} = 
  \sqrt{1\over m\left({\sf X}^T{\sf X}\right)^{-1}_{ii}}\;.
\end{equation}

\subsubsection{Signal matrix for maximum performance}

We now derive the properties of the signal matrix $\sf X$ that
provides maximum performance, i.e.\ minimizes the sum of squares of
the rows of $\sf Y$ under the condition that ${\sf X}{\sf Y} = {\sf 1}$.
Using Eq.~\ref{equ:keller_OptPolEff}, del Toro Iniesta and Collados
(2000) showed that the squares of the maximum possible
polarimetric efficiencies are given by
\begin{equation}
\epsilon_{{\rm max},i}^2 = {1\over m}\sum_{j=1}^m {\sf X}_{ji}^2
                         ={1\over m}\left({\sf X}^T {\sf X}\right)_{ii}\;.
\end{equation}
Hence the maximum possible efficiency is given by the sum of squares
of the elements of the {\em columns} of the signal matrix $\sf X$,
normalized with the number of measurements.  

For ${\sf Y}{\sf X} = {\sf 1}$ to hold (remember that we have only
considered the diagonal elements of this relation so far), we must
also require that
\begin{equation}
\sum_{j=1}^m{\sf Y}_{ij}{\sf X}_{jk} = \delta_{ik}\;,
\end{equation}
where $\delta_{ik}$ is equal to 0 unless $i=k$ where it takes on the
value 1.  Inserting
\begin{equation}
{\sf Y}_{ij} = {{\sf X_{ji}}\over \sum_{k=1}^m {\sf X}_{ki}^2}\;.
\label{equ:OptInverseElement}
\end{equation}
into the above equation and multiplying both sides of the equation
with the denominator of the left side yields
\begin{equation}
\sum_{j=1}^m{\sf X}_{ji}{\sf X}_{jk} = 
  \delta_{ik}\sum_{j=1}^m{\sf X}_{jk}^2\;.
\end{equation}
For $i=k$, the equation is obviously correct.  For $i\ne k$, the right
side is zero and we conclude that ${\sf X}^T{\sf X}$ has to be
diagonal for a polarimeter to achieve the maximum performance.

A polarimeter that achieves its maximum possible polarimetric
efficiency therefore has a signal matrix $\sf X$ such that
\begin{equation}
{\sf X}^T{\sf X} = m\left(
\begin{array}{cccc}
\epsilon_{{\rm max},1}^2 & 0 & 0 & 0 \\
0 & \epsilon_{{\rm max},2}^2 & 0 & 0 \\
0 & 0 & \epsilon_{{\rm max},3}^2 & 0 \\
0 & 0 & 0 & \epsilon_{{\rm max},4}^2 \\
\end{array}
\right)\;.
\end{equation}

\subsubsection{Generation of maximum performance signal matrices}

We now go beyond the work of del Toro Iniesta and Collados (2000) to
provide more insight into the properties of the signal matrix $\sf X$
for a polarimeter with maximum efficiency.  Since each row of the
signal matrix $\sf X$ is the first row of a Mueller matrix, it obeys
the inequality
\begin{equation}
{\sf X}_{i1}^2 \ge \sum_{j=2}^4 {\sf X}_{ij}^2\;,
\end{equation}
where the equal sign applies if there are no depolarizing elements
between the source and the detector.  This is a direct consequence of
the properties of a Mueller matrix.

The optimum polarimeter has an intensity efficiency of $\epsilon_1=1$.
The other three efficiencies have to obey $\sum_{i=2}^4 \epsilon_i^2
\le 1$.  This implies that the optimum efficiency for measuring
polarized Stokes components with equal efficiencies is given by
${1\over\sqrt{n}}$ where $n$ is the number of polarized Stokes
parameters that are measured.  If all Stokes parameters are measured,
the maximum efficiency is ${1\over\sqrt{3}} \approx 0.577$, if only
two are measured, the maximum efficiency is ${1\over\sqrt{2}} \approx
0.707$, and for a single polarized Stokes parameter, the maximum
efficiency is obviously 1.

In a polarimeter with maximum efficiency, the elements of the first
column of the signal matrix $\sf X$ are equal to 1.  Since the columns
of the signal matrix need to be orthogonal, and the scalar product of
the first column with any of the other columns corresponds to the sum
over all elements of the other product, we conclude that
\begin{equation}
\sum_{j=1}^m {\sf X}_{jk}=0,\ k=2...4\;.
\end{equation}

For a polarimeter reaching maximum efficiency, there will be no
depolarizing elements, and we have
\begin{equation}
\sum_{k=2}^4 {\sf X}_{jk}^2=1,\ j=1...m\;,
\end{equation}
since the rows of $\sf X$ correspond to the first rows of Mueller matrices.
We can thus consider each row of the signal matrix as a point on
the Poincar\'e sphere.  In the following, we will show that by
maximizing the average distance squared between these points (on the
Poincar\'e sphere), we obtain a signal matrix that provides maximum
polarimetric efficiency.

To maximize the separation between points on the Poincar\'e sphere, we
maximize the $m$ functions
\begin{equation}
\sum_{j\ne i}^m\sum_{k=2}^4\left({\sf X}_{ik}-{\sf X}_{jk}\right)^2 -
  \alpha_i\left(\sum_{k=2}^4{\sf X}_{ik}^2 -1\right)\;,
\end{equation}
where $\alpha_i$ are Lagrange multipliers.  These equations are valid
independent of the number of polarized components that are considered
since the elements ${\sf X}_{jk}$ are zero for polarized
components of the Stokes vector that we do not want to measure.

By setting the derivatives of this function with respect to ${\sf
  X}_{ik}$ to zero, we obtain
\begin{equation}
\left(m-\alpha_i\right){\sf X}_{ik} = \sum_{j=1}^m{\sf X}_{jk}\;.
\label{equ:OptColumnSum}
\end{equation}
By squaring both sides of this equation, summing over $k$, and
remembering that $\sum_{k=2}^4{\sf X}_{ik}^2=1$, we obtain
\begin{equation}
\left(m-\alpha_i\right)^2 = 
\sum_{k=2}^4\left(\sum_{j=1}^m{\sf X}_{jk}\right)^2\;.
\end{equation}
Since the right side is independent of $i$, we conclude that all
Lagrange multipliers take on the same value, i.e.\ $\alpha_i=\alpha$.
Using this relation in Eq.~\ref{equ:OptColumnSum} and summing over all
$i$, we obtain
\begin{equation}
\left(m-\alpha\right)\sum_{i=1}^m{\sf X}_{ik} = m\sum_{j=1}^m{\sf X}_{jk}\;.
\end{equation}
Because $\alpha=0$ is not a valid Lagrange multiplier, the only way to
fulfill this equation is the requirement that the sums of columns of
$\sf X$ for $k=2...4$ vanish, i.e.\
\begin{equation}
\sum_{j=1}^m{\sf X}_{jk} = 0\;.
\end{equation}
In other words, the points have to be distributed on the Poincar\'e
sphere in such a way that their center of gravity is always at the
origin of the sphere.  This is the same requirement as for a
polarimeter with maximum efficiency.

We conclude with calculating the distance $\Delta$ between points on the
Poincar\'e sphere.  The average distance squared between one point and
all the other points is given by
\begin{equation}
\Delta^2 = {1\over m-1}\sum_{j\ne i}^m\sum_{k=2}^4
  \left({\sf X}_{ik} - {\sf X}_{jk}\right)^2\;,
\end{equation}
which reduces to
\begin{equation}
\Delta^2 = {2\left(m-1\right)+2\over m-1}\;.
\end{equation}
We finally obtain
\begin{equation}
\Delta = \sqrt{2+{2\over m-1}}\;.
\end{equation}
For $m=2$ we obtain $\Delta=2$, which corresponds to opposite sides of
the Poincar\'e sphere.  For $m=3$ we obtain $\Delta=\sqrt{3}$, which
corresponds to the side length of a triangle whose plane includes the
origin of the Poincar\'e sphere.  For $m=4$ we obtain
$\Delta=\sqrt{8\over 3}$, which corresponds to the side length of a
tetrahedron inside the Poincar\'e sphere.  Note, that $\Delta$ only
corresponds to the distance between points if the distance between
{\em all} points is the same, which is the case for these examples.
It should not be surprising that this average distance is independent
of the number of polarized Stokes components that are considered,
since at no point in our derivation did we make any assumptions about
the number of components that we want to measure.  

Unfortunately, maximizing the average square distance of points is a
necessary but not a sufficient criteria for generating a signal matrix
corresponding to a polarimeter with maximum performance.  The
following signal matrix
\begin{equation}
\sf X = \left(
\begin{array}{rrrr}
1 &  x &  x &  x \\
1 &  x &  x &  x \\
1 & -x & -x & -x \\
1 & -x & -x & -x \\
\end{array}
\right)
\end{equation}
also has an average distance of $\sqrt{8\over3}$, but ${\sf X}^T{\sf
X}$ is obviously not diagonal.

\section{Optimum Calibration}

Once a polarization analysis system has been designed, we need to
determine the optimum way to calibrate it, i.e.\ to experimentally
determine the signal matrix $\sf X$.  As it turns out, much of the
previously derived results can also be applied to find the optimum
calibration approach, i.e.\ based on the measured signal matrix $\sf
X$, we need to determine the optimum matrix $\sf Y$.
To determine all 16 elements of $\sf X$, we need to make (at least)
16 measurements of the signal $\mathbf S_i,i=1..m$, which corresponds to
(at least) 4 different input Stokes vectors $\mathbf I^c_i,i=1..m$ with
(hopefully) known properties.

Based on the approach by Azzam et al.\ (1988), we group the 4
calibration input Stokes vectors $\mathbf I_i$ and the corresponding
signal vectors $\mathbf S_i$ into 4 by $m$ matrices.  We can then write
\begin{equation}
{\sf S} = {\sf X} {\sf I}^c\;,
\end{equation}
with
\begin{eqnarray}
{\sf S} &=& \left({\mathbf S}_1 {\mathbf S}_2 ... {\mathbf S}_m\right)\;, \\
{\sf I}^c &=& \left({\mathbf I}_1^c {\mathbf I}_2^c ... {\mathbf I}_m^c\right)\;.
\end{eqnarray}
The signal matrix $\sf X$ is given by ${\sf X} = {\sf S} {\sf J}$
with ${\sf J} {\sf I}^c = {\sf 1}$.  We need to choose ${\sf I}^c$ such as
to minimize the error in $\sf X$, given errors in the measurements
$\sf S$.  This is the same problem that we faced when
optimizing the polarimetric efficiency, and we can indeed apply the
same reasoning.  The calibration input Stokes vectors ${\mathbf I}_i^c$ for
maximum calibration accuracy should therefore obey the same relations
as the rows of the signal matrix $\sf X$ of a polarimeter with
maximum performance.

For a polarimeter that only measures one polarized component (e.g.\ a
circular polarimeter), we would use the two corresponding orthogonal
polarization states.  For a system that measures two polarized
components (e.g.\ a linear polarimeter), we would use three
calibration Stokes vectors whose points are the corners of an
equilateral triangle.  For the linear polarimeter, we could use a
rotating linear polarizer positioned $60\deg$ apart.  For a
vector-polarimeter, we would choose the corners of a tetrahedron as
suggested by Azzam et al.\ (1988).  

In practice, however, one has to take many more measurements so that
the properties of the non-ideal polarization calibration optics can
also be determined.

\section{Polarization Error Budget}

\subsection{Classical Error Budgets}

Error budgets are a classical tool in systems engineering to derive
requirements for the individual parts of a system such that the system
as a whole meets the requirements while minimizing the total
complexity and/or cost.  A typical example is the wavefront aberration
in an optical system that contains many optical components.  An
academic example is shown in Tab.~\ref{tab:keller_ErrorBudgetOptical}.

\begin{table}[!ht]
\caption{Error budget example for the optical
quality (80\% encircled energy in arcsec) of a 
ground-based telescope with an instrument.
\label{tab:keller_ErrorBudgetOptical}}
\smallskip
\begin{center}
{\small
\begin{tabular}{lllrrr}
\tableline
\noalign{\smallskip}
level 1 item&level 2 item&level 3 item & level 1 & level 2 & level 3 \\
\noalign{\smallskip}
\tableline
\noalign{\smallskip}
atmosphere & & & 0.50 & & \\
telescope & & & 0.25 & \\
 & primary mirror & & & 0.17 & \\
 & & mirror polishing & & & 0.10 \\
 & & mirror support  & & & 0.10 \\
 & & thermal distortion & & & 0.10 \\ 
 & secondary mirror & & & 0.17 & \\
 & & mirror polishing & & & 0.10 \\
 & & mirror support  & & & 0.10 \\
 & & thermal distortion & & & 0.10 \\ 
instrument & & & 0.25 & & \\
total & & & 0.61 & & \\
\noalign{\smallskip}
\tableline
\end{tabular}
}
\end{center}
\end{table}

An error budget can have several levels since parts can again be
looked at as a combinations of smaller parts.  Since the overall error
is estimated from the combination of many sources, mistakes in the
estimates of errors of individual components tend to average out.
Furthermore, only the overall error has a requirement attached to it,
and the individual errors of each level of the error budget can be
allocated in different ways.  This error allocation is an iterative
process where one tends to minimize the complexity and cost of the
system while focusing on the main contributors to the system error,
which are quickly identifiable in the error budget.

For an error budget to make sense, one needs to figure out how to add
errors.  In the case of aberrations of an optical system, the
wavefront errors of individual elements are not correlated, and one
can simply assume that the final error is given by the square root of
the sum of errors squared of the individual elements, the so-called
{\em root sum of squares} (RSS).  For a polarimeter, the issue is much
more complicated since 1) polarization is a vector quantity and not a
scalar; 2) retarders and polarizers affect the polarization in a major
way and not in a minor way as required for RSS to make sense; and 3)
errors of very different nature are combined non-linearly, such as the
instrumental polarization of the optical system that feeds the
polarimeter and the non-linearity of the light detector system (Keller
1996).

Another difficulty comes from the fact that a polarimeter is, in most
cases, calibrated experimentally.  This means that some (but not all)
errors can be drastically reduced, and one should only include the
error that remain after the calibration.  This is somewhat analogous to
wavefront aberration error budgets for optics that include active
and/or adaptive optical elements.

\begin{table}[!ht]
\caption{Schematic polarization error budget for a polarimeter.
\label{tab:keller_ErrorBudgetPolarized}}
\smallskip
\begin{center}
{\small
\begin{tabular}{lll}
\tableline
\noalign{\smallskip}
level 1 item&level 2 item&level 3 item \\
\noalign{\smallskip}
\tableline
\noalign{\smallskip}
source variation & & \\
atmosphere & & \\
telescope & & \\
polarimeter & & \\
 & polarizer & \\
 & detector system & \\
 & & undetected bias variation \\
 & & nonlinearity \\
calibration & & \\
 & polarizer & \\
 & retarder & \\
 & & positioning repeatability \\
 & & temperature change of retarder \\
data reduction & & \\
\noalign{\smallskip}
\tableline
\end{tabular}
}
\end{center}
\end{table}

Because of all these difficulties, {\em polarization} error budgets
have historically not been used in the design of polarimeters.  First
attempts have been described by Boger et al.\ (2003), but this does
not go further than a list of potential errors.
Table~\ref{tab:keller_ErrorBudgetPolarized} shows several potential
error sources in polarimeters, separated into three levels.

\subsection{Errors in Mueller Matrices}

If an optical element in a polarimeter can be described by a Mueller
matrix, then any small error associated with this element alone can be
approximated by a linear change in the Mueller matrix elements.  For
example, a Mueller matrix ${\sf M}\left(\alpha,\beta\right)$ describes
an optical element with two parameters $\alpha$ and $\beta$, e.g.\ a
retarder with retardation and fast axis orientation.  A Taylor
approximation for the Mueller matrix with scalar errors $\delta\alpha$
and $\delta\beta$ in the respective scalar parameters $\alpha$ and
$\beta$ can be written as
\begin{equation}
{\sf M}\left(\alpha+\delta\alpha,\beta+\delta\beta\right) \approx
{\sf M}\left(\alpha,\beta\right) + {\sf m}_\alpha\cdot\delta\alpha
                                 + {\sf m}_\beta\cdot\delta\beta
\label{equ:keller_taylormatrix}
\end{equation}
For some parameters of some elements it can be that the first-order
Taylor expansion is insufficient and that the second order has to be
considered.  For simplicity, we will assume in the following that the
first-order expansion is adequate.  Instead of working with specific
errors $\delta\alpha$ and $\delta\beta$, we need to look at
Eq.~\ref{equ:keller_taylormatrix} in a statistical sense such that
$\delta\alpha$ and $\delta\beta$ correspond to characteristic values
of the magnitude of the expected error, e.g.\ the standard deviation
for a zero-mean Gaussian distribution or the extreme values of a
uniform distribution.  The matrices ${\sf m}_{\alpha,\beta}$ can then
be interpreted as normalized standard deviations of the elements of
the Mueller matrix $\sf M$.

For a uniform distribution of errors in $\alpha$ over $\pm\Delta$ we
can calculate the variance of the Mueller matrix with respect to
errors in $\alpha$ as
\begin{equation}
{\sf M}_\alpha^2\left(\Delta\right) = \int_{-\Delta}^{+\Delta}\left(
    {\sf M}\left(\alpha+\epsilon,\beta\right)-
    {\sf M}\left(\alpha,\beta\right)\right)^2\delta\epsilon\;,
\end{equation}
where the $^2$ does not indicate a matrix multiplication but the
square of the individual matrix elements.  The normalized standard 
deviation then becomes
\begin{equation}
{\sf m}_\alpha = \lim_{\Delta\Rightarrow 0}\frac{\partial}{\partial\Delta}
                 {\sf M}_\alpha\left(\Delta\right)\;.
\end{equation}
Hence, the Mueller matrix with errors can be written as
\begin{equation}
{\sf M}\left(\alpha,\beta\right) \pm {\sf m}_\alpha\cdot\delta\alpha
                                 \pm {\sf m}_\beta\cdot\delta\beta\;.
\end{equation}

As an example, we show a linear retarder with fast axis angle $\theta$
and retardance $\phi$. The corresponding Mueller matrix is {\tiny
\begin{equation}
\left(
\begin{array}{llll}
 1 & 0 & 0 & 0 \\
 0 & \cos ^2(2 \theta )+\cos (\phi ) \sin ^2(2 \theta ) & \cos (2 \theta ) \sin (2 \theta )-\cos (2 \theta ) \cos (\phi ) \sin (2 \theta ) & \sin (2 \theta ) \sin (\phi ) \\
 0 & \cos (2 \theta ) \sin (2 \theta )-\cos (2 \theta ) \cos (\phi ) \sin (2 \theta ) & \cos (\phi ) \cos ^2(2 \theta )+\sin ^2(2 \theta ) & -\cos (2 \theta ) \sin (\phi ) \\
 0 & -\sin (2 \theta ) \sin (\phi ) & \cos (2 \theta ) \sin (\phi ) & \cos (\phi )
\end{array}
\right)
\end{equation}
}
For a uniform error distribution in retardance $\phi$ we obtain the normalized standard deviation matrix with respect to $\phi$ as
{\small
\begin{equation}
\left(
\begin{array}{llll}
 0 & 0 & 0 & 0 \\
 0 & \frac{\sqrt{\sin ^4(2 \theta ) \sin ^2(\phi )}}{\sqrt{3}} & \frac{\sqrt{\sin ^2(4 \theta ) \sin ^2(\phi )}}{2 \sqrt{3}} & \frac{\sqrt{\cos ^2(\phi ) \sin ^2(2 \theta
   )}}{\sqrt{3}} \\
 0 & \frac{\sqrt{\sin ^2(4 \theta ) \sin ^2(\phi )}}{2 \sqrt{3}} & \frac{\sqrt{\cos ^4(2 \theta ) \sin ^2(\phi )}}{\sqrt{3}} & \frac{\sqrt{\cos ^2(2 \theta ) \cos ^2(\phi
   )}}{\sqrt{3}} \\
 0 & \frac{\sqrt{\cos ^2(\phi ) \sin ^2(2 \theta )}}{\sqrt{3}} & \frac{\sqrt{\cos ^2(2 \theta ) \cos ^2(\phi )}}{\sqrt{3}} & \frac{\sqrt{\sin ^2(\phi )}}{\sqrt{3}}
\end{array}
\right)\;.
\end{equation}
}

\subsection{Statistical Distribution of Parameter Errors}

In the best case scenario, the statistical distribution for the random
errors is known from a large number of components with identical
requirements.  However, this is often not the case and one has to make
assumptions.  While a Gaussian distribution might be a natural choice
for certain alignment errors, it is often too optimistic an
assumption, in particular for manufacturing errors.  When components
are manufactured, the manufacturer often stops the processing once the
component meets the specifications.  It is therefore likely that the
component will have properties close to the maximum errors, which is a
very different distribution from a normal distribution where the
parameter is most likely to be at the specified value.  A uniform
distribution of the errors is therefore often a better choice.

\subsection{Combining Mueller Matrix Errors}

Stenflo (1994) showed that weakly polarizing Mueller matrices ${\sf
M}_i$ can be written as ${\sf E} + {\sf m}_i$ where $\sf E$ is the
unity matrix and $m_{i,jk}\ll 1$ and that the product of such
matrices can be approximated by their sum.  Therefore, RSS can be
applied to weakly polarizing and retarding elements.

This result can be generalized to strongly polarizing and retarding
elements written as ${\sf M}_i + {\sf m}_i$ and again $m_{i,jk}\ll 1$.
However, ${\sf M}_i$ is not the unity matrix, and all matrix elements
$M_{i,jk}$ can be of order unity.  The product of such matrices can
then be approximated according to
\begin{equation}
  \prod_{i=1}^{n}\left(\sf{M}_i+\sf{m}_i\right)  \approx
  \prod_{i=1}^{n}\sf{M}_i + 
    \sum_{i=1}^n\left(\prod_{j=1}^{i-1}\sf{M}_j\right)\sf{m}_i
    \left(\prod_{j=i+1}^n\sf {M}_j\right)
\end{equation}
The resulting product becomes the sum of the product of error-free
matrices and all the Mueller matrix errors transformed using the ideal
Mueller matrices of elements before and after the current element.
Because the transformed errors are additive, we can RSS these
transformed Mueller matrix errors and use a classic error budget
approach to estimate the contribution of individual errors to the
overall system.

\section{Outlook}

To make these tools useful, systematic errors that cannot be expressed
in terms of Mueller matrices must be included, and calibration and
data reduction errors must be added.  A library of normalized standard
deviation matrices for common polarimetric components and their
respective parameters must be collected.

The polarimetry error budget described here neglects the coupled
effect of simultaneous errors in all design parameters.  A Monte Carlo
simulation avoids this drawback by considering simultaneous errors.
However, such end-to-end simulations require substantial efforts and
do not provide a direct insight into the main error contributors and
how their effects can be balanced.

Only the application of these proposed techniques to several real
instruments will show whether this is indeed a useful tool for
designing polarimeters.


\end{document}